\documentclass[aps,onecolumn,superscriptaddress,notitlepage]{revtex4-1}
\usepackage{epsfig,amsmath,amssymb,amsfonts,physics,graphicx,color,calc,epstopdf}

\usepackage[T1]{fontenc}

\usepackage{xcolor,hyperref}
\hypersetup{
   colorlinks,
   linkcolor={blue!50!black},%{red!80!black},
   citecolor={blue!50!black},
   urlcolor={blue!80!black}
}
\let\a=\alpha \let\b=\beta  
\let\e=\varepsilon   
\let\l=\lambda    
\let\s=\sigma \let\t=\tau

 \let\r=\rho

 \def\HH{{\cal H}}
 
\def\LL{{\cal L}}  \def\OO{{\cal O}}
\def\DD{{\cal D}}
 \def\SS{{\cal S}}
  
\def\ZZ{{\cal Z}}

\def\Im{{\rm Im}\,}

\def\de{\mathrm{d}}

\newcommand{\beq}{\begin{equation}} 
\newcommand{\eeq}{\end{equation}}
\newcommand{\ba}{\begin{eqnarray}}
\newcommand{\ea}{\end{eqnarray}}

\begin{document}

\title{Field theory for zero temperature  soft anharmonic spin glasses in a field}

\author{Pierfrancesco Urbani}
\affiliation{Universit\'e Paris-Saclay, CNRS, CEA, Institut de physique th\'eorique, 91191, Gif-sur-Yvette, France}

 \begin{abstract}
We introduce a finite dimensional anharmonic soft spin glass in a field and show how it allows the construction a field theory at zero temperature and the corresponding loop expansion.  The mean field level of the model coincides with a recently introduced fully connected model, the KHGPS model, and it has a spin glass transition in a field at zero temperature driven by the appearance of pseudogapped non-linear excitations. We analyze the zero temperature limit of the theory and the behavior of the bare masses and couplings on approaching the mean field zero temperature critical point. Focusing on the so called replicon sector of the field theory, we show that the bare mass corresponding to fluctuations in this sector is strictly positive at the transition in a certain region of control parameter space. At the same time the two relevant cubic coupling constants $g_1$ and $g_2$ show a non-analytic behavior in their bare values: approaching the critical point at zero temperature,  $g_1\to \infty$ while $g_2\propto T$ with a prefactor diverging at the transition. 
Along the same lines we also develop the field theory to study the density of states of the model in finite dimension. We show that in the mean field limit the density of states converges to the one of the KHGPS model. However the construction allows a treatment of finite dimensional effects in perturbation theory.
\end{abstract}

\maketitle

\section{Introduction}
The fate of spin glass transition in finite dimension has been at the heart of a big debate in the last 40 years.
The main issue is the nature of the spin glass phase itself.
Two main scenarios have been proposed. The first one originates in the mean field theory of spin glasses and it is based on the replica approach. According to this view the spin glass phase is described by replica symmetry breaking which implies many pure states organized in an ultrametric way \cite{MPV87}.
Another point of view, the droplet theory, based on scaling arguments of local excitations, suggests that the spin glass phase is a disguised ferromagnet and there is no space for ultrametricity and complex landscape of pure states, characteristic of the replica symmetry breaking scenario \cite{fisher1987absence, fisher1988equilibrium}. 

Despite a huge research effort to clarify this issue, no final answer has been found. 
Recently, this problem has become central because while finite dimensional spin glasses have a small number of applications, glasses and amorphous solids are more widespread from this point of view. In \cite{KPUZ13} it has been proposed, using the mean field limit of infinite spatial dimensions, that glasses, when forced to be at the bottom of their free energy landscape, therefore both at low enough temperature or high pressure in the case of colloidal glasses, may undergo to a so called Gardner transition \cite{SimpleGlasses2020}. In this case, a glass, cooled out-of-equilibrium, goes from a replica symmetric phase to a full replica symmetry breaking phase. The relation with spin glasses is that the nature of the Gardner transition is the same as the spin glass transition in a field \cite{BU15} and the same is true for the low temperature phase, at least at the mean field level. This is due to very simple symmetry considerations. In the high temperature glass the system has no other symmetry than the replica one. Therefore on lowering the temperature or at high pressures, the only symmetry that can break is only the replica one.

The droplet and the replica symmetry breaking scenarios differ for a crucial aspect: the former predicts that one cannot have any spin glass transition in an external field while the latter predicts a transition also in the presence an external field.

Perturbative renormalization group computations to discriminate between these two pictures are inconclusive \cite{bray1980renormalisation, pimentel2002spin, BU15} when they are carried out in the region where perturbation theory can be safely used \footnote{See also \cite{charbonneau2017nontrivial} for a semi-perturbative study which returns a fixed point in a non-perturbative region of coupling constants.}. 
The main reason for this is that the perturbative renormalization group has no perturbative fixed point below the upper critical dimension where the Gaussian fixed point looses its stability. 
One possibility is that the transition is not destroyed but is driven by a non-perturbative fixed point inaccessible from perturbative schemes. Another possibility is that the transition changes nature from being continuous to discontinuous.
As way to circumvent this problem, it has been suggested \cite{parisi2012replica, angelini2022unexpected} to consider directly the theory at zero temperature and applied field and test the stability of this zero temperature critical point when finite dimensional fluctuations are switched on. This route has been applied to the Random Field Ising model \cite{parisi1979random}, where it is known that the zero temperature critical point exists and its attractive from a renormalization point of view, but one needs to clarify its critical nature (when, for example, dimensional reduction applies \cite{tissier2011supersymmetry, fytas2019evidence, kaviraj2020random}).
However in order to build on such a program one needs to have a simple mean field starting point with a genuine spin glass transition in a field at zero temperature.
This has been found recently in \cite{bouchbinder2020low} and it has been named the KHGPS model. The model has been analyzed in the RSB phase in \cite{folena2021marginal} where it has been shown that for sufficiently strong interactions and external magnetic field, the zero temperature spin glass transition is driven by the appearance of pseudogapped non-linear excitations which are responsible for a chaotic behavior of the energy landscape under applied field implying system spanning equilibrium avalanches. The purpose of this work is to build a genuinely finite dimensional model which at the mean field level coincides with the KHGPS model and that allows to attack finite dimensional effects both at the zero temperature spin glass transition in a field and in the replica symmetric phase.
{In order to do that we follow the very same route as in \cite{moore20121}} and introduce a vector model and use it to perform a loop expansion. We will then develop the theory around its mean field limit and consider the first non-trivial cubic coupling constants which generate the loop expansion. We show that their bare value diverges on approaching the mean field critical point in the interesting region where the spin glass transition is driven by the appearance of pseudogapped non-linear excitations. 
Finally we will use the very same model to compute the field theory for the resolvent function from which we can extract the density of states at zero temperature in the region where the external field is sufficiently high (the replica symmetric phase). 
We show that the mean field limit of this construction gives back the equations for the resolvent found in  \cite{bouchbinder2020low}.

\section{Definition of the model}
The construction of a field theory from a mean field model can be carried out in many ways. The simplest one is to consider the degrees of freedom on a $d$-dimensional lattice and let them interact on a finte ranged interaction matrix \cite{amit2005field}. If the range of interaction is sent to infinity (as in the Kac limit for example) one gets a mean field model. When the interaction range is finite instead one has a bona fide finite dimensional model. This route has been used in \cite{temesvari2002generic} to construct a generic replica field theory for Ising spin glasses in finite dimensions. However in the case of  \cite{temesvari2002generic} the initial starting point, the mean field level of the theory corresponding to the SK model, does not allow for a spin glass transition at zero temperature and finite external field. Here we want to build on the KHGPS model which has a bona fide spin glass transition in a field at zero temperature and define a field theory that can be considered directly at $T=0$ coming from the replica symmetric side of the mean field critical point. 
Instead of proceeding as in \cite{temesvari2002generic} we consider a slightly different route which we believe it could be useful for numerical simulation purposes \footnote{Simulating long range models is a bit more complicated computationally than finitely connected models.}
We generalize the KHGPS model to a non-linear vector model and we compute the field theory as a loop expansion induced around the mean field limit of infinite size of the vector variables. 

We consider the following model. The degrees of freedom are real variables $x_{i\a}$. The index $i=1,\ldots, N$ runs on a $d$-dimensional lattice. The index $\alpha=1,\ldots, M$ is a vectorial index.
The Hamiltonian of the model is given by
\beq
H[\underline x] = \frac{J}{\sqrt{2d M}}\sum_{i<j} c_{ij} \sum_{\a \b}J_{ij}^{\a\b} x_{i\a} x_{j\b} + \sum_{i=1}^N \sum_{\a=1}^M v(x_{i\a},k_{i\a}) 
\label{MModel}
\eeq
where we have indicated with $c_{ij}$ the adjacency matrix of the corresponding $d$-dimensional lattice. We note that in what will follow we can easily consider matrices $c$ that do not describe $d$-dimensional lattices but are more generic.  We will also assume that there are no self interactions so that $c_{ii}=0$ for all $i=1,\ldots, N$.
The random couplings $J_{ij}^{\a\b}$ are Gaussian random numbers with zero mean and unit variance.
The local potentials are also random and given by
\beq
v(x_{i\a},k_{i\a}) = \frac{k_{i\a}}{2} x_{i\a}^2+ \frac{1}{4!} x_{i\a}^4 - h_{\rm ext} x_{i\a}
\eeq
and the elastic constants $k_{i\a}$ are random $i.i.d$ variables distributed according to a fixed probability distribution $p(k)$ defined in the interval $[k_{\min},k_{\max}]$ and we will assume that $k_{\min}>0$ which has the interesting phenomenology \cite{bouchbinder2020low}. To fix the ideas we will consider $p(k)=1/(k_{\rm min}-k_{\rm max})$ but other forms of $p(k)$ are not supposed to produce any substantial difference \footnote{apart changes of non-universal properties of the phase diagram like the precise location of the transition lines and non-universal prefactors.} as long as $k_{\rm min}>0$ and $p(k_{\rm min})>0$ \footnote{See \cite{rainone2021mean} and \cite{lacroix2022counting} for a similar model where the $p(k)$ is gapless}.
We also added the external field $h_{\rm ext}$ that breaks the ${\mathbb Z}_2$ symmetry $x_{i\a}\to -{x_{i\a}}$.
The construction of this model follows a route that is similar to what is done in second order phase transitions to compute critical exponents at fixed spatial dimension by enlarging the space of degrees of freedom to a vector space, see \cite{bray1974self, parisi1980field}. Note however that in the $\phi^4$-field theory of \cite{bray1974self, parisi1980field} one generalizes the model to an $O(N)$-invariant model, while here we will not add any microscopic global (or gauge) symmetry (at fixed couplings). 
{This idea of using vector models to perform the loop expansion in spin glasses was used in \cite{moore20121} building on a $O(M)$ spin glass model. Here we follow the very same construction.}
Moreover, we note that vectorial models for $O(M)$ spins have been used recently in \cite{franz2022delocalization, franz2022linear} to model the density of states in zero temperature amorphous solids. The model in Eq.\eqref{MModel} differs from \cite{franz2022delocalization, franz2022linear} because of two points: (i) we consider real (unbounded) spins subjected to a random anharmonic (quartic) local potential which is essential for the physical behavior close to the mean field spin glass transition at zero temperature, and (ii) we include, {as in \cite{moore20121}}, the adjacency matrix $c$ and use the $M\to \infty$ limit  as a way to generate a field theory expansion at fixed and finite spatial dimension.

For $d\to \infty$ and $M=1$ the model reduces to the KHGPS model as much as in the same limit the Edwards-Anderson model \cite{edwards1975theory} gives the Sherrington-Kirkpatrick model \cite{sherrington1975solvable}. We will consider the more interesting case where $d<\infty$ and we will show that for $M\to \infty$ we get the same phase diagram of the KHGPS model. This implies that the model has a zero temperature spin glass transition at finite external field in the $M\to \infty$ limit. An expansion around this limit allows the construction of a loop expansion which should give the properties of the model in finite dimension and finite $M$ and we compute the bare masses and coupling constants at zero temperature of the corresponding field theory.

\section{Construction of the field theory}
{We follow exactly the same strategy as in \cite{moore20121}.}
We start by defining the partition function $Z$ as
\beq
Z = \int_{-\infty}^\infty \left[\prod_\a^M \prod_i^N\de  x_{i\a}\right]\, e^{-\hat \beta H[\underline x]}
\eeq
where we have indicated with $\hat \b = 1/T$ the inverse temperature. 
We are interested in the free energy of the model
\beq
{\rm f} = -\lim_{N\to \infty}\frac 1{\hat \beta N} \overline{\ln Z}
\eeq
which can be obtained using the replica trick
\beq
{\rm f} = -\lim_{N\to \infty}\frac 1{\hat \beta N} \overline{\ln Z} = - \lim_{N\to \infty} \frac{1}{\hat \b N}\partial_n \overline{Z^n}\:.
\eeq
The overline denotes the average over all sources of disorder, meaning both $J_{ij}^{\a\b}$s and the $k_{i\a}$s.
For $n$ real replicas we have
\beq
\overline{Z^{n}} =\overline{ \int_{-\infty}^\infty \left[\prod_a^n\prod_\a^M \prod_i^N\de  x_{i\a}^{(a)}\right] \exp[-\hat\beta \sum_a H[\underline x^{(a)}]]}
\eeq
and we have denoted by $\underline x^{(a)}=\{x_{i\a}^{(a)}\}_{i=1,\ldots, N; \a=1,\ldots, M}$.
The average over the disorder can be performed explicitly.
There are two relevant terms.
The first and most important one is the average over the spin glass couplings $J$s.
We need to average the term
\beq
\begin{split}
&I_1= \overline{\exp\left[-\frac{\hat \beta J}{\sqrt{2dM}} \sum_{i<j} \sum_{\a \b}c_{ij} J_{ij}^{\a\b}\sum_{a}^n x_{i\a}^{(a)}x_{j\b}^{(a)}  \right]} \propto\exp\left[\frac{{\hat\beta}^2J^2}{8d M } \sum_{ij} \sum_{ab} \left(\sum_\a x_{i\a}^{(a)}x_{i\a}^{(b)}\right)c_{ij}\left(\sum_\b x_{j\b}^{(a)}x_{j\b}^{(b)}\right) \right]
\end{split}
\eeq
and we have neglected irrelevant proportionality factors.
We can then use a Hubbard-Stratonovich transformation so that the last term can be rewritten as
\beq
\begin{split}
I_1=\int {\cal D Q} \exp\left[-\frac{{\hat \beta}^2 M J^2 d}{2}\sum_{ij}\sum_{ab} q_i^{ab}[c^{-1}]_{ij}q_j^{ab}+ \frac{\hat\beta^2J^2}{2}\sum_i\sum_{ab} q_i^{ab} \sum_{\a} x_{i\a}^{(a)}x_{i\a}^{(b)}\right]
\end{split}
\eeq
and with ${\cal DQ}$ we have indicated the measure over all variables $q_i^{ab}$. Note that we are integrating over all variables $q_i^{ab}$ but it is clear that the action has a natural symmetry $q_i^{ab} = q_i^{ba}$ which is unbroken.
The average over the local elastic constants factorizes on the single sites and therefore one can write the replicated partition function as
\beq
\overline{Z^n} = \int {\cal D Q} \exp[S_n[{\cal Q}]]
\eeq
where the action is 
\beq
S_n[{\cal Q}] = -\frac{{\hat \beta}^2 M J^2 d}{2}\sum_{ij}\sum_{ab} q_i^{ab}[c^{-1}]_{ij}q_j^{ab} + M \sum_i \ln {\cal Z}_i
\label{full_action}
\eeq
and the local partition functions $Z_i$ define the local \emph{impurity} problems
\beq
\begin{split}
{\cal Z}_i &= \int_{k_{\min}}^{k_{\max}} \de p(k) \int_{-\infty}^\infty \left[ \prod_a \de x_a\right] \exp\left[ -\sum_a^n\hat \beta v(x_a,k) +\frac{\hat \beta^2 J^2}{2}\sum_{ab} q_i^{ab} x_a x_b\right]\:.
\end{split} 
\eeq
The action in Eq.~\eqref{full_action} defines the field theory we want to look at. 
It is clear that at finite $d$ and $M$ one has a complicated theory to solve. We will show in the following that taking the $M\to \infty$ limit allows to get a saddle point solution which describes a mean field theory which has a genuine spin glass transition at zero temperature and finite external field with interesting properties.

\subsection{The $M\to \infty$ limit and mean field theory}
We now consider the $M\to \infty$ limit.
The integral over $\cal Q$ can then be evaluated by saddle point. The corresponding equations are
\beq
q_i^{ab} = \frac 1{2d} \sum_j c_{ij} \langle x_a x_b\rangle_j
\label{local_SP}
\eeq
where the brackets are averages over the Boltzmann measure over the impurity problems given by
\beq
\begin{split}
\langle \OO(x_1,\ldots,x_n) \rangle_i &= \frac{1}{{\cal Z}_i} \int_{k_{\min}}^{k_{\max}} \de p(k) \int_{-\infty}^\infty \left[\prod_a^n \de x_a\right]\OO(x_1,\ldots,x_n) \exp \left[ -\hat \beta \sum_{a=1}^nv(x_a,k) +\frac{\hat \beta^2 J^2}{2}\sum_{ab} q_i^{ab} x_a x_b\right]
\end{split}
\label{imp_av}
\eeq
and we have denoted by $\OO$ a generic function.
The interpretation of Eq.~\eqref{local_SP} is clear. One has an overlap order parameter $q_i^{ab}$ which is local on each site.
Mean field theory gives it as an average of the neighboring overlaps.
It is clear that translational invariance imposes $q_{i}^{ab} = q_\star^{ab}$ given by
\beq
q_\star^{ab}  = \langle x_a x_b\rangle\:.
\eeq
This saddle point equation coincides with the one derived in \cite{bouchbinder2020low, folena2021marginal}.

We now summarize the properties of the saddle point solution \cite{bouchbinder2020low, folena2021marginal}.
For any $J$, the model has a spin glass transition  in the temperature-external field plane for sufficiently low temperature and sufficiently small external field \cite{folena2021marginal}. Furthermore at zero temperature there is a spin glass transition when the field is lowered below a critical value \cite{bouchbinder2020low}.
The properties of this critical point depend on the strength of $J$. In particular if $J$ is sufficiently small, the zero temperature spin glass transition is driven by the divergence of the spin glass susceptibility. This happens when at the transition point the ground state develops an abundance of soft linear excitations (in particular the density of states at this point goes as $\rho(\l)\sim \l^{1/2}$). 

If $J$ is large enough instead, the corresponding critical point is not driven by the growth of the spin glass susceptibility (for strictly zero temperature) \cite{bouchbinder2020low} since the density of states has a higher pseudogap $\rho(\l)\sim \l^{3/2}$. In this case the phase transition is driven by the appearance of pseudogapped non-linear excitations \cite{folena2021marginal} whose properties at the mean field level are described by the full-replica-symmetry-breaking scheme of Parisi \cite{MPV87}. We will consider this second case which is more interesting from the physical viewpoint and comment later of the more standard case where also at $T=0$ the spin glass susceptibility diverges.

In the replica symmetric phase, the saddle point solution is parametrized by
\beq
\begin{split}
q_d&=\langle x_a^2\rangle\\
q&=\langle x_a x_b\rangle \ \ \ \ \ a\neq b \:.
\end{split}
\label{RS_q}
\eeq
It is convenient to rewrite the saddle point equation for the diagonal term $q_d$ and use the variable $\chi$ defined as 
\beq
\chi =\beta \left[\langle x_a^2\rangle -\langle x_a x_b\rangle_{a\neq b}\right] =\beta (q_d-q)\:.
\label{RS_chi}
\eeq
In the zero temperature limit $\chi>0$.
Plugging this ansatz inside Eq.~\eqref{full_action} we get that the free energy of the impurity problems, for $n\to 0$ reduces to 
\beq
\ln {\cal Z}_i \simeq n \int_{k_{\rm min}}^{k_{\rm max}} \de p(k)\int_{-\infty}^\infty Dz \ln \int \de x \exp\left[-\hat \beta \hat v (x|k,z)\right]
\eeq
where the effective potential is
\beq
\hat v(x|k,z)= \frac{1}{2}(k-J^2\chi)x^2 + \frac{x^4}{4!} -(h_{\rm ext} + J\sqrt{q} z)x
\label{eff_potential}
\eeq
and we have used that the saddle point does not depend on the site index.
The measure $Dz$ indicates a Gaussian measure with zero mean and unit variance $Dz=\de z\,e^{-z^2/2}/\sqrt{2\pi}$.
In this way the parameter $\chi$ becomes the linear magnetic susceptibility defined as
\beq
\chi =\beta \overline{\langle x^2 \rangle_c}
\eeq
and the overline denotes the average with respect to $k$ and $z$ while the brakets denote a Boltzmann average with Hamiltonian given by the effective potential in Eq.~\eqref{eff_potential} and we have denoted with the subscript $c$ the connected correlation function.

For large enough $J$, the transition to RSB at $T=0$ appears when \cite{bouchbinder2020low}
\beq
k_{\rm min} - J^2\chi = 0
\label{transizione_omega4}
\eeq
which corresponds to the condition for the appearance of an effective single site potential having a double well shape for a certain region in the $(k,z)$ plane.
In the RSB phase, the model has a fullRSB solution and the order parameter becomes a continuous function $q(y)$ with $y\in [0,1]$ \cite{MPV87} whose shape is such that $\de q/\de y >0$ in a certain range of $y$. The fullRSB solution renormalizes the distribution of the (cavity) fields whose distribution is encoded in $h_{\rm ext} + J\sqrt{q} z$. In particular the random variable $z$ is not Gaussian anymore but it has a distribution dictated by the RSB equations \cite{folena2021marginal}. 
Interestingly, this distribution depends on $k$ and while it is regular for $k\geq J^2\chi$, for $k\in[k_{\min},J^2\chi]$ it has a linear pseudogap centered in $z=-h_{\rm ext}/(J\sqrt q)$ \cite{folena2021marginal}. This corresponds to a suppression of the density of effective potential shapes having two degenerate wells. This also implies that in the RSB phase the evolution of the energy landscape is chaotic under applied field. At a given external field, one has a ground state surrounded by many metastable minima. Changing infinitesimally the field, this ground state becomes a metastable state and some previously metastable minimum, maybe far in phase space, becomes the ground state. This induces an equilibrium avalanche. It is important to note that both minima do not disappear and could be followed in principle \footnote{This is for example easily seen in models where all $k_{i\a}$ are the same and negative. In this case the spectrum in the ground state is even gapped as it cane be shown by extending the results of \cite{folena2021marginal}, implying that minima can be followed at linear level.}. The evolution of the ground state is therefore fully non-perturbative, something which is precisely taken into account by the RSB solution.

We will now consider the loop expansion coming from the unbroken replica symmetric phase and characterize the replica field theory from this side of the transition. The study of the broken phase is much harder due to the fact that mean field theory predicts a continuous spectrum of vanishing masses which are directly related to the marginally stable character of the RSB solution \cite{goltsev1983stability, de1983eigenvalues, crisanti2010sherrington, crisanti2011stability}.
 
\subsection{Setting up the loop expansion}
We now consider the $1/M$ expansion which will generate a loop expansion.
We consider fluctuations away from the saddle point:
\beq
q_i^{ab} = q_\star^{ab} + \delta q_i^{ab}\:.
\eeq
Note that we will always consider symmetric matrices so that we need to consider only $a\leq b$ and enforce whenever is needed that $\delta q_i^{ab} = \delta q_i^{ba} $. Following closely \cite{temesvari2002generic} we will denote by $\sum_{(a,b)} = \sum_{a\leq b}$ while $\sum_{ab}$ denotes a sum over all couples $a,b$.
Expanding the replicated action of Eq.~\eqref{full_action} around the saddle point we get
\beq
\begin{split}
\delta S_n[\delta {\cal Q}] \equiv S_n[{\cal Q_*} + \delta {\cal Q}] - S_n[{\cal Q_*}]&= -\frac{{\hat \beta}^2 M J^2 d}{2}\sum_{ij} \sum_{ab}\delta q_i^{ab}\, [c^{-1}]_{ij}\, \delta q_j^{ab} +  \frac M2 \sum_{i} \sum_{(a,b), (c,d)} \left.\frac{\partial^2 {\cal Z}_i}{\partial q_i^{ab} \partial q_i^{cd}}\right|_{{\cal Q_*}} \delta q_i^{ab}\delta q_i^{cd}\\
&+\frac M{3!} \sum_{(a,b),(c,d), (e,f)}\left.\frac{\partial^3 {\cal Z}_i}{\partial q_i^{ab} \partial q_i^{cd} \partial q_i^{ef}}\right|_{{\cal Q_*}} \delta q_i^{ab}\delta q_i^{cd}\delta q_i^{ef}+\ldots
\end{split}
\eeq
Translational invariance implies that the coefficients of the expansion are independent on the site index $i$. 
Therefore the field theory becomes
\beq
\overline{Z^n} = e^{S_n[{\cal Q}_\star]}\int {\cal D} \delta{\cal Q}\exp\left[\delta S_n[\delta {\cal Q}]\right]
\eeq
Since we are interested in studying the theory close to zero temperature we consider the following rescaling
\beq
\hat\beta \sqrt M \delta q_i^{ab} = \hat{\delta q}_i^{ab}
\eeq
so that
\beq
\begin{split}
\overline{Z^n} &\propto \int {\cal D} \hat{\delta{\cal Q}}\exp\left[ -\frac{J^2 d}{2}\sum_{ij} \sum_{ab}\hat{\delta q}_i^{ab}\, [c^{-1}]_{ij}\, \hat{\delta q}_j^{ab} +  \frac 1{2\hat \b^2} \sum_{i} \sum_{(a,b), (c,d)} \left.\frac{\partial^2 {\cal Z}_i}{\partial q_i^{ab} \partial q_i^{cd}}\right|_{{\cal Q_*}} \hat{\delta q}_i^{ab}\hat{\delta q}_i^{cd}\right.\\
&\left.+\frac 1{3! \sqrt M \hat \b^3} \sum_{(a,b),(c,d), (e,f)}\left.\frac{\partial^3 {\cal Z}_i}{\partial q_i^{ab} \partial q_i^{cd} \partial q_i^{ef}}\right|_{{\cal Q_*}} \hat{\delta q}_i^{ab}\hat{\delta q}_i^{cd}\hat{\delta q}_i^{ef}+\ldots\right]
\end{split}
\label{exp_gen}
\eeq
Therefore the $1/M$ perturbation theory generates clearly a loop expansion.
At this point we need to compute the bare couplings and masses, meaning the coefficients of the expansion of the impurtity problems.
This can be done in full generality, even in the broken replica symmetry phase. However the loop expansion in the RSB phase is complicated. We limit ourselves to look at the theory in the replica symmetric phase where these coefficients can be easily computed. This program was performed in \cite{temesvari2002generic} by expanding around the SK model at finite temperature close to the spin glass transition. Here we will do the same expanding around the KHGPS model.
We now underline that the fluctuations $\hat{\delta q}_i^{ab}$ have different components. 
One can have both diagonal terms $\hat{\delta q}_i^{aa}$ and off diagonal fluctuations $\hat{\delta q}_i^{a\neq b}$.
A classification of the types of fluctuations that are relevant around the replica symmetric saddle point can be made systematic by looking at the masses of the theory enforced by the quadratic term of the expansion in Eq.~\eqref{exp_gen}. It is well known that one has three eigenvalues, denoted as the replicon, the anomalous and the longitudinal \cite{de1978stability, temesvari2002generic}. We do not repeat the analysis here. A pedagogical derivation can be found in \cite{nishimori2001statistical} \footnote{Note that the analysis of \cite{nishimori2001statistical} involves a case in which one has fluctuations of the off-diagonal matrix elements $q_{a\neq b}$ and of an additional set of order parameters that are the magnetizations $m_a$. However the analysis in \cite{nishimori2001statistical} relies on purely algebraic structure. Therefore one can exchange the role of $m_a$ with the diagonal part $q_{aa}$ with the overlap matrix in our case.}. While we could redo the analysis in full generality, in the following we will focus only on the replicon sector. This is what, at the mean field level, drives the transition at finite temperature and therefore we want to study the fate of these fluctuations in the zero temperature limit. A systematic study of all the other sectors is left for future works.

\section{The replicon sector}
The replicon sector is defined by fluctuations that are only off diagonal (namely ${\delta q}_i^{aa}=0$) and that satisfy the following sum rule
\beq
\sum_{a=1}^n \hat{\delta q}_i^{ab}=\sum_{b=1}^n \hat{\delta q}_i^{ab} = 0\:.
\eeq
One can then re-organize the perturbative expansion on this subspace \cite{temesvari2002generic}. We will consider the mass term of the field theory and then the first non-trivial cubic interaction terms.

\subsection{The mass term}
The mass term is defined by
\beq
\begin{split}
-\frac{ J^2 d}{2}\sum_{ij} \sum_{ab}\hat{\delta q}_i^{ab}\, [c^{-1}]_{ij}\, \hat{\delta q}_j^{ab} +  \frac {\hat \b^2}2 \sum_{i} \sum_{(a,b), (c,d)} \left.\frac{\partial^2 {\cal Z}_i}{\partial q_i^{ab} \partial q_i^{cd}}\right|_{{\cal Q_*}} \hat{\delta q}_i^{ab}\hat{\delta q}_i^{cd} =\frac 12 \sum_{\underline p} \hat{\delta q}^{ab}(\underline p) M_{ab;cd}(\underline p)\hat{\delta q}^{cd}(-\underline p)
\end{split}
\eeq
where we have used a summation over the Fourier modes being $\underline p$ a $d$-dimensional vector conjugated to the lattice matrix $c$.
The mass term of the replicon sector can be computed from standard algebraic methods, see \cite{nishimori2001statistical,temesvari2002generic} and at long wavelengths it is given by
\beq
m_{R}(p) \propto |\underline p|^2 + r_R 
\eeq
where we have neglected order one proportionality factors which are irrelevant for this analysis.
The value of $r_R$ is then given by
\beq
r_R \propto 1 - J^2 \hat\beta^2\overline{\langle x^2\rangle_c^2 }
\eeq
where we have neglected strictly finite proportionality factors and where the brackets now indicate the Boltzmann measure
\beq
\begin{split}
\langle \cdot \rangle &= \frac 1{Z(k,z)}\int_{-\infty}^\infty \de x\, \cdot \, e^{-\hat \beta \hat v(x|k,z)}\\
Z(k,z)&=\int_{-\infty}^\infty \de x\, e^{-\hat \beta \hat v(x|k,z)}
\end{split}
\eeq
and the overline is the average over $k$, with probabability density $p(k)$, and over the Gaussian variable $z$ which has zero mean and unit variance.
According to \cite{bouchbinder2020low}, the zero temperature spin glass transition may or may not be accompained by a vanishing replicon mass. If the external field $h_{\rm ext}$ is small, one has that the transition is signaled by $r_R\to 0$. The more interesting situation is instead at large external field where the mean field transition point has $r_R>0$ at the transition.
We will show that at zero temperature, the case in which $r_{R}\to 0$ corresponds to the typical case considered for example in \cite{FPUZ15}. The bare coupling constants in the replicon sector are small (or they can be made small by large $M$) and therefore the starting point can be perturbative. Conversely, when the bare mass $r_R>0$ at the mean field critical point, 
the bare couplings diverge.

\subsection{Cubic coupling constants}
Having artificially fronzen the fluctuations of the local overlaps $\hat{\delta q}_i^{ab}$ on the sectors other than the replicon one, we are left with two coupling constants $g_1$ and $g_2$ \cite{temesvari2002generic}. 
If we denote by $\phi_i^{ab}$ the components of $\hat{\delta q}_i^{ab}$ in the replicon sector, the corresponding cubic interaction terms are
\beq
\delta S^{(3)} = \frac{1}{\sqrt M} \sum_i^N\left[g_1  \Tr [\phi_i]^3 +\frac 12 g_2 \sum_{ab} \left(\phi_i^{ab}\right)^3\right]
\eeq
The expressions of the bare cubic coupling constants are given by
\beq
\begin{split}
g_1 &\propto  {\hat \b^3}\overline{\langle x^2\rangle_c^3}\\
g_2 &\propto {\hat \b^3}\overline{\langle x^3\rangle_c^2}
\end{split}
\eeq
and we have neglected $\OO(1)$ proportionality constants.
We can now study the behavior of these coupling constants approaching the critical point at zero temperature. We underline that in what follows we will focus on the case where the $T=0$ critical point has a finite spin glass susceptibility ($r_R>0$).
We can rewrite the two coupling constants by introducing the effective free energy as
\beq
f(k,z) = \ln \int_{-\infty}^\infty \de x \exp[-\hat v(x|k,z)]
\eeq
For $\hat \b\to \infty$ the integral is dominated by a saddle point $x^*(z,k)$ whose equation is
\beq
(k-J^2\chi)x^*+ \frac{{x^*}^3}{6}-\left(h_{\rm ext}+J\sqrt{q} z\right)=0
\label{eq_x**}
\eeq
where $\chi$ and $q$ are determined by Eqs.~(\ref{RS_q},\ref{RS_chi}).
Solving for $x^*$ we get
\beq
\begin{split}
g_1\propto&\overline{\left[\frac{1}{k-J^2\chi + {x^*}^2/2}\right]^3}\\
\lim_{\hat \beta \to \infty}\hat \beta g_2\propto& \overline{\frac{{x^*}^2}{\left(k-J^2\chi + {x^*}^2/2\right)^6}}\:.
\end{split}
\eeq 
and again the overline stands for the average over $z$ and $k$.
We will now compute the right hand side of these expressions. We consider $J$ fixed and call $H_c$ the critical value of the external field where Eq.~\eqref{transizione_omega4} is satisfied. For $h_{\rm ext}>H_c$ the denominator in these expressions does not vanish in the domain of both random variables $k$ and $z$. Therefore we have that these expressions are finite far from the critical point. This implies that $g_2\propto T$.
We now compute how these quantities behave when $h_{\rm ext}\to H_c^+$. We define the random variable
\beq
m=k-J^2\chi + \frac{{x^*}^2}{2}
\eeq
whose support is defined in $m\in[k_{\rm min}-J^2\chi, k_{\rm max}-J^2\chi]$. Therefore when $h_{\rm ext}\to H_c^+$ we get that the left edge of the support of $m$ touches zero. We now show that both $g_1$ and $\hat \beta g_2$ diverge as $h_{\rm ext}\to H_c^+$. Instead of computing the explicit dependence on $h_{\rm ext}$ we parametrize the distance from the critical point implicitly with $m_{\rm min}=k_{\rm min}-J^2\chi$ which goes to zero as $h_{\rm ext} \to H_c^+$. All parameters, namely $\chi$ and $q$ are regular and finite at the critical point so we can neglect their explicit dependence on $h_{\rm ext}$.
We start the analysis by computing $g_1$. Using the same technique as in the appendix of \cite{bouchbinder2020low} we get
\beq
g_1\propto \int_{m_{\rm min}}^{m_{\rm max}} \frac{\de m}{\Delta m} \int_{-\infty}^\infty \frac{\de x}{\sqrt{2\pi J^2 q}} \frac{1}{\left(m+\frac{x^2}{2}\right)^2} e^{A[m,x]}
\eeq 
where 
\beq
\begin{split}
A[m,x]&=-\frac{1}{2J^2q}\left(\frac{x^3}{6} + m x-h_{\rm ext}\right)^2\\
\Delta m&= m_{\rm max} - m_{\rm min}\\
m_{\rm max} &= k_{\rm max} - J^2\chi \:.
\end{split}
\eeq
It is easy to show by integration by parts that we can isolate the most diverging contribution to $g_1$ when $m_{\rm min}\to 0$ which is given by
\beq
g_1\sim m_{\rm min}^{-1/2} 
\eeq
The very same analysis can be carried out for $\hat \b g_2$ which gives
\beq
\hat \b g_2\sim m_{\rm min}^{-5/2}\:.
\eeq
Therefore at the bare critical point, both $g_1$ and $\hat \beta g_2$ diverge.

\subsubsection{The coupling constants in the case where the transition is driven by $r_R=0$.}
In the region of control parameter space where the transition is driven by the vanishing of the replicon mass, the same computation for $g_1$ and $\hat \b g_2$ gives back two bare coupling constants that are finite at the transition point.
Therefore, for large enough values of $M$ both coupling constants are small and one start with a genuinely perturbative field theory which is precisely the same as \cite{bray1980renormalisation}. Note however that in the present case $g_2\sim T\to 0$ which must be compared to what happens at zero external field where $g_2$ is identically equal to zero by ${\mathbb Z}_2$ symmetry. 

In the scenario in which the transition becomes non-perturbative below the upper critical dimension, one may have that the flow of the RG goes to a non-perturbative fixed point which coincides with the one controlling the transition in the case in which the bare replicon mass does not vanish at the mean field level. This may be a possibility since it is clear that the mean field mechanism giving rise to the divergence of the spin glass susceptibility at zero temperature (psuedogapped density of states going as $\l^{1/2}$  with completely delocalized eigenvectors) is fragile in finite dimensions.

\section{The density of states}
We consider now the density of states of the ground state in the model.
We are interested in computing the density of eigenvalues of the Hessian matrix defined as
\beq
\HH_{i\a,j\b} = \frac{J}{\sqrt{2dM}}c_{ij}J_{ij}^{\a\b} + \delta_{ij}\delta_{\a\b}v''(x_{i\a}^*,k_{i\a})
\label{hessian_form}
\eeq
The density of eigenvalues can be computed using the formula \cite{edwards1976eigenvalue}
\beq
\r(\l)=\lim_{N\to \infty}\lim_{\e\to 0^+}\frac{1}{\pi}\Im G(\l-i\e)
\eeq
where the resolvent $G(z)$ is given by
\beq
G(z)=\frac 1{NM} \overline{\mathrm{Tr} \left[z{\bf 1} - \HH \right]^{-1}}
\eeq
and we have indicated with an overline the average with respect to the stiffnesses $k_{i\a}$ and the couplings $J$s.
We note that the ground state vector $x^*$ depends on the couplings and on the stiffnesses and therefore it is random.
We first consider $M$ fixed as much as the dimension of the lattice $d$. Then we will take the $M\to \infty$ limit and shown that the resolvent satisfies the same equation of the KHGPS model of \cite{bouchbinder2020low}. 

\subsection{The replica field theory for the resolvent}
In order to compute the resolvent, we follow the pioneering route proposed by Edwards and Jones \cite{edwards1976eigenvalue} and write it as a two point function for a Gaussian theory as follows
\beq
\begin{split}
G(z) &= \frac{1}{NM} \lim_{m\to 0} \left(-\frac{2}{m}\right)\frac{\de}{\de z} \lim_{\hat \b\to \infty}{\mathbb E}_{k,J} \frac{1}{Z} \int \de \underline x\, e^{-\hat \beta H[\underline x]} \\
&\times \int_{-\infty}^\infty \left[\prod_{\s=1}^m\prod_{i=1}^N\prod_{\a=1}^M\frac{\de \phi_{i\a}^{(\s)}}{\sqrt{2\pi/i}}\right]\exp\left[-\frac{i}{2} \sum_{\s=1}^m\sum_{i\a,j\b}\phi_{i\a}^{(\s)}\left(\left(z - v''(x_{i\a},k_{i\a})\right)\delta_{ij}\delta_{\a\b} - \frac{J}{\sqrt{2dM}} c_{ij}J_{ij}^{\a\b}\right)\phi_{j\b}^{(\s)}\right]\:.
\end{split}
\eeq
We note now that we need to introduce an additional set of replicas in order to treat the partition function $Z$. 
Therefore we get
\beq
\begin{split}
G(z) &= \frac{1}{NM} \lim_{n\to 0}  \lim_{m\to 0}\left(-\frac{2}{m}\right)\frac{\de}{\de z} \lim_{\hat \b\to \infty}{\mathbb E}_{k,J}  \int \DD (\phi,x)\\
&\times \exp\left[-\sum_{a=1}^n\hat \beta H[\underline x^{(a)}]-\frac{i}{2} \sum_{a=1}^n\sum_{\s=1}^m\sum_{i\a,j\b}\phi_{i\a}^{(a,\s)}\left(\left(z - v''(x_{i\a}^{(a)},k_{i\a})\right)\delta_{ij}\delta_{\a\b} - \frac{J}{\sqrt{2dM}} c_{ij}J_{ij}^{\a\b}\right)\phi_{j\b}^{(a,\s)}\right]
\end{split}
\eeq
where we have used a shorthand for the integration measure:
\beq
\DD (\phi,x)  = \prod_{i=1}^N\prod_{\a=1}^M \prod_{a=1}^n\de x_{i\a}^{(a)} \prod_{\s=1}^m \frac{\de \phi_{i\a}^{(a,\s)}}{\sqrt{2\pi/i}}
\eeq
and for all variables the domain of integration is the real line.
We can now average over the disorder.
The terms that depend on the interaction matrix $J$ are
\beq
I_J={\mathbb E}_J\exp\left[-\frac{\hat \beta J}{\sqrt{2Md}} \sum_{i<j} \sum_{\a \b}c_{ij} J_{ij}^{\a\b}\sum_{a=1}^n x_{i\a}^{(a)}x_{j\b}^{(a)} +i  \sum_{\s=1}^m\sum_{a=1}^n \sum_{i<j}\sum_{\a\b}\frac{J}{\sqrt{2dM}} c_{ij}J_{ij}^{\a\b}\phi_{i\a}^{(a,\s)}\phi_{j\b}^{(a,\s)}\right]\:.
\eeq
Averaging over the  matrix $J$ we get the following expression (neglecting irrelevant proportionality constants)
\beq
\begin{split}
&I_J\propto\exp\left[\frac{{\hat\beta}^2J^2}{8d M } \sum_{ij} \sum_{ab} \left(\sum_\a x_{i\a}^{a}x_{i\a}^{b}\right)c_{ij}\left(\sum_\b x_{j\b}^{a}x_{j\b}^{b}\right) -\frac{J^2}{8dM}\sum_{ij}\sum_{(a,\s)(b,\t)} \left(\sum_{\a}\phi_{i\a}^{(a,\s)}\phi_{i\a}^{(b,\t)}\right)c_{ij}\left(\sum_{\b}\phi_{j\b}^{(a,\s)}\phi_{j\b}^{(b,\t)}\right)\right.\\
&\left. -\frac{i J^2\hat \b}{4dM}\sum_{ij}\sum_{\a\b} \sum_{(a,\s) b} \left(\sum_\a\phi_{i\a}^{(a,\s)}x_{i\a}^{(b)}\right)c_{ij}\left(\sum_\b \phi_{j\b}^{(a,\s)}x_{j\b}^{(b)}\right)\right]\\
&= \int \DD P \DD Q \DD T \exp\left[-\frac{{\hat \beta}^2 M J^2 d}{2}\sum_{ij}\sum_{ab} q_i^{ab}[c^{-1}]_{ij}q_j^{ab}+ \frac{\hat\beta^2J^2}{2}\sum_i\sum_{ab} q_i^{ab} \sum_{\a} x_{i\a}^{(a)}x_{i\a}^{(b)}\right]\\
&\times \exp\left[-\frac{M J^2 d}{2}\sum_{ij}\sum_{(a,\s)(b,\t)} p_i^{a\s,b\t}[c^{-1}]_{ij}p_j^{a\s,b\t}+ i\frac{J^2}{2}\sum_i\sum_{(a,\s)(b,\t)} p_i^{a\s,b\t} \sum_{\a} \phi_{i\a}^{(a,\s)}\phi_{i\a}^{(b,\t)}\right]\\
&\times \exp\left[i\frac{M J^2 d}{2}\sum_{ij}\sum_{(a,\s)b} t_i^{a\s,b}[c^{-1}]_{ij}t_j^{a\s,b}+ i\frac{J^2}{2}\sum_i\sum_{(a,\s)b} t_i^{a\s,b} \sum_{\a} \phi_{i\a}^{(a,\s)}x_{i\a}^{(b)}\right]\:.
\end{split}
\eeq
Finally the disorder in the elastic constants $k$s is factorized on the sites. Therefore the resolvent can be written as
\beq
G(z) = \frac{1}{NM} \lim_{n\to 0}  \lim_{m\to 0}\left(-\frac{2}{m}\right) \frac{\de}{\de z}\lim_{\hat \b\to \infty} \int \DD P \DD Q \DD T \exp\left[ \SS \left(P,Q,T\right)\right]
\eeq
where the action is given by
\beq
\begin{split}
\SS(P,Q,T) &= -\frac{{\hat \beta}^2 M J^2 d}{2}\sum_{ij}\sum_{ab} q_i^{ab}[c^{-1}]_{ij}q_j^{ab}-\frac{M J^2 d}{2}\sum_{ij}\sum_{(a,\s)(b,\t)} p_i^{a\s,b\t}[c^{-1}]_{ij}p_j^{a\s,b\t}\\
& +i\frac{M J^2 d}{2}\sum_{ij}\sum_{(a,\s)b} t_i^{a\s,b}[c^{-1}]_{ij}t_j^{a\s,b} + M \sum_{i} \ln\ZZ_{i}\:.
\end{split}
\eeq
The impurity problem now is given by
\beq
\begin{split}
\ZZ_i &= {\mathbb E}_k\int \left[\prod_{a\s} \frac{\de \phi^{(a,\s)}}{\sqrt{2\pi/i}}\de x^{(a)} \right]e^{\LL} \\
\LL & = \frac{\hat\beta^2J^2}{2}\sum_{ab} q_i^{ab} x^{(a)}x^{(b)}+ i\frac{J^2}{2}\sum_{(a,\s)(b,\t)} p_i^{a\s,b\t} \phi^{(a,\s)}\phi^{(b,\t)}+ i\frac{J^2}{2}\sum_{(a,\s)b} t_i^{a\s,b} \phi^{(a,\s)}x^{(b)} \\
&-\sum_{a=1}^n\left[ \hat \b v(x^{(a)},k) + \frac i2 \sum_{\s=1}^m (z-v''(x^{(a)},k))\left(\phi^{(a,\s)}\right)^2
\right]\:.\end{split}
\eeq
This concludes the construction of the field theory. It is important to note that the action for the field $T$ is symmetric under the ${\mathbb Z}_2$.

\subsection{The density of states in the $M\to \infty$ limit}
In the $M\to \infty$ limit one can evaluate the resolvent using a saddle point.
The action for $T$ is symmetric under $T\to -T$ and therefore the saddle point solution is $t_i^{(a,\s)b} = 0$. Furthermore we will suppose to be in the replica symmetric phase so that the parametrization of the matrix $q$ is as in Eq.~\eqref{RS_q}. The corresponding ansatz for the matrix $P $ is 
\beq
p_i^{(a,\s)(b,\t)} = \Pi(z)\delta_{ab}\delta_{\s\t}
\eeq
where we have used translational invariance.
The saddle point equation for $q$, given the $m\to 0$ limit from the outset, coincides with the one studied in Eq.~\eqref{RS_q}.
The equation for $\Pi(z)$, in the $n\to 0$ limit is instead given by
\beq
\Pi(z) = {\mathbb E}_{k,h}\frac{1}{z-v''(x^*(h,k),k)-J^2\Pi(z)}
\label{ppp}
\eeq
where $x^*$ is the solution of Eq.~\eqref{eq_x**} being $h$ a Gaussian random variable with zero mean and unit variance.
Using the saddle point equations we can easily show that $G(z)\equiv \Pi(z)$.
Indeed Eq.~\eqref{ppp} coincides with the one derived in \cite{bouchbinder2020low} for the KHGPS model and it predicts that when $J$ is sufficiently large, the replica symmetric phase has a gapped spectrum which becomes gapless at the zero temperature spin glass transition. If the external field is sufficiently large, at the transition point the spectrum behaves as $\rho(\l)\sim \l^{3/2}$ implying that the spin glass susceptibility is finite, while for small external field, the spectrum goes as $\rho(\l)\sim \l^{1/2}$ implying a divergence of the spin glass susceptibility.

\subsection{The expansion around the saddle point}
It is clear that as is was done for the field theory itself, a $1/M$ expansion is also possible for the density of states. This is left for future work. 
However we would like to discuss what one could expect from the outcome of such a computation. 
It must be understood that $1/M$ corrections encode for correlation between the diagonal part and the off-diagonal part of the Hessian matrix in \eqref{hessian_form}. In the replica symmetric region, at high enough external field, the $M\to\infty$ density of states has a gap and the natural question is if this gap is filled beyond the $M\to \infty$ limit. We know from numerical simulations in similar models, see for example \cite{das2020robustness}, that one may expect that in any finite dimension, the density of states has a quartic tail at low frequencies. Therefore one may conjecture that the same happens for this model at any finite $M$ and $d$ and this must be tested in simulations. If this is the case one may wonder if the gap in the $M\to \infty$ density of states is filled already by the first $1/M$ correction. This possibility must be explored by a careful computation.

Finally we would like to point out that the same $1/M$ expansion can be used to compute in finite dimension the \emph{forces} defined as
\beq
f_{i\a} = \frac{J}{\sqrt{2dM}}\sum_{j}\sum_\b c_{ij}J_{ij}^{\a\b} x_{j\b}\:.
\eeq
These quantities have been shown to be very important, see \cite{das2020robustness, folena2021marginal}. In the $M\to \infty$ case they are Gaussian random variables in the replica symmetric phase (while in the RSB phase they behave in a non-trivial way \cite{folena2021marginal}). In finite dimension they deviate from the Gaussian behavior and how to get them has been conjectured in \cite{das2020robustness} to be a difficult problem. Our proposal is to use the $1/M$ expansion as a tool to compute non-Gaussian corrections in finite dimension. This is left for future work.

\section{Discussion and conclusions}
We have considered a model for anharmonic vectorial spins as a tool to construct a field theory for the spin glass phase in a field at zero temperature. 
Furthermore we have shown that in the region of control parameter space where the mean field limit of the theory gives a zero temperature spin glass transition in a field driven by the appearance of non-linear excitations, the cubic coupling constants of the field theory diverge at the bare level. This may imply that the transition (if any) becomes non-perturbative. Another possibility is that this hints to a first order behavior at $T=0$. However this would imply a finite temperature critical point on the de Almeida-Thouless transition line where the transition goes from first order to a continuous transition on increasing temperature. This seems to be ruled out by recent numerical simulations \cite{fernandez2022numerical} and mean field theory gives a continuous transition at all finite temperature critical point \cite{folena2021marginal}. Note that if one approaches the transition from the finite temperature mean field spin glass transition \cite{folena2021marginal} one finds a different behavior where the spin glass susceptibility is always divergent at arbitrarily small temperature and the cubic coupling constants $g_1$ and $g_2$ diverge with a ratio that converges numerically close to $g_2/g_1\sim1/2$ \cite{folena2021marginal} a picture which is qualitatively coherent with numerical simulations \cite{fernandez2022numerical}.  
Finally it is clear that since at the mean field level the $T=0$ transition is driven by the appearance of non-linear excitations, a way to characterize the critical point is by looking at equilibrium avalanches. If the zero temperature spin glass transition in a field is stable in finite dimension, one should observe scale free equilibrium avalanches appearing on lowering the field beyond the critical point, while in the replica symmetric phase the distribution of avalanche size is expected to have an exponential cutoff for large sizes.

Last but not least, we believe that the model defined in Eq.~\eqref{MModel} may be useful for numerical simulations. For supposedly moderate large values of $M$ and large enough external field, one should be able to find the ground state with good accuracy using message passing algorithms \cite{mezard2009information} in finite dimension by closing the belief propagation equations on Gaussian distributions \cite{antenucci2019approximate}, and use the corresponding outcome as a starting point for a simulated annealing to refine the solution. This would help to contrast theoretical predictions with numerical simulations.

Finally it is clear that the very same construction applies when $c$ describes interactions decaying as a power-law in the distance between spins (the classic example is the one of dipolar interactions) and therefore it is amenable to study the case in which the model describes the interaction of degrees of freedom mediated by an elastic kernel. The same construction can be used to study finite dimensional corrections to the Gradient descent dynamics when applied to the path integral describing the dynamical partition function. The same consideration holds when one switches on quantum fluctuations. The model in Eq.\eqref{MModel} can be canonically quantized by adding a kinetic term of the form $\sum_{i\a} p^{2}_{i\a}/2$ and promote positions and momenta to have the canonical commutation relations $[x_{i\a},p_{j\b}] = i\hbar \delta_{ij}\delta_{\a\b}$. In all these cases, it is already very interesting to study the replica symmetric phase at high external magnetic field.   
 
 \paragraph*{Acknowledgments ---} 
The author is grateful to G. Folena, T. Maimbourg and J.-M. Luck for useful discussions. This work is supported by ``Investissements d'Avenir'' LabEx PALM (ANR-10-LABX-0039-PALM).

%\bibliography{refs.bib}

\end{document}